\documentclass[12pt]{report}
\input BoxedEPS
\SetRokickiEPSFSpecial  
\HideDisplacementBoxes
\usepackage{color,amsmath,amssymb}
\def\thebibliography#1{\leftline{\large\bf References}\list
  {[\arabic{enumi}]}{\settowidth\labelwidth{[#1]}\leftmargin\labelwidth
\advance\leftmargin\labelsep
\usecounter{enumi}}
\def\newblock{\hskip .11em plus .33em minus .07em}
\sloppy\clubpenalty4000\widowpenalty4000}

\newcommand{\vek}[1]{\mathbf{#1}}

\renewcommand{\arctan}{\tan^{-1}}

\newcommand{\pvalt}{\raise0.15ex\hbox{-}\mkern-11.5mu\int}
\newcommand{\be}{\begin{equation}}
\newcommand{\ee}{\end{equation}}
\newcommand{\bea}{\begin{eqnarray}}
\newcommand{\eea}{\end{eqnarray}}
\newcommand{\ben}{\begin{enumerate}}
\newcommand{\een}{\end{enumerate}}
\newcommand{\bit}{\begin{itemize}}
\newcommand{\eit}{\end{itemize}}

\newcommand{\lab}[1]{\label{#1}}

\newcommand{\half}{\frac{1}{2}} 

\newcommand{\p}{\partial}

\renewcommand{\ln}{\,\mbox{ln}\,}

\newcommand{\de}{\delta}
\newcommand{\De}{\Delta}

\newcommand{\eps}{\epsilon}

 \newcommand{\la}{\lambda}
\newcommand{\La}{\Lambda}

\newcommand{\si}{\sigma}

\newcommand{\beq}{\begin{equation}}
\newcommand{\eeq}{\end{equation}}
\newcommand{\ba}{\begin{array}}
\newcommand{\ea}{\end{array}}

\newcommand\hide[1]{} 

\begin{document}
    
\begin{center}
{\large \bf
The Dirichlet Casimir Problem}
\end{center}

\centerline{
N.~Graham$^{\rm a}$,
R.L.~Jaffe$^{\rm b}$, V.~Khemani$^{\rm b}$,
M.~Quandt$^{\rm c}$, O.~Schr\"oder$^{\rm b}$,
H.~Weigel$^{\rm  c,d}$
}

\parbox[t]{15cm}{
\begin{center}
{~}\\$^{\rm a}$Department of Physics,
Middlebury College \\
Middlebury, VT 05753 \\~\\
$^{\rm b}$Center for Theoretical Physics,
Laboratory for Nuclear Science\\ and Department of Physics,
Massachusetts Institute of Technology\\ Cambridge, Massachusetts 02139 \\~\\
$^{\rm c}$Institute for Theoretical Physics, T\"ubingen University\\
D-72076 T\"ubingen, Germany\\~\\
$^{\rm d}$Fachbereich Physik, Siegen University\\
D-57068 Siegen, Germany\\
{~} \\
{  \qquad MIT-CTP-3413 \qquad UNITU-HEP-14/2003 \\
hep-th/0309130}
\end{center}
}
 
\centerline{\large\bf Abstract}

{\small\noindent Casimir forces are conventionally computed by
analyzing the effects of boundary conditions on a fluctuating quantum
field.  Although this analysis provides a clean and calculationally
tractable idealization, it does not always accurately capture the
characteristics of real materials, which cannot constrain the modes of
the fluctuating field at all energies.  We study the vacuum
polarization energy of renormalizable, continuum quantum field theory
in the presence of a background field, designed to impose a Dirichlet
boundary condition in a particular limit.  We show that in two and
three space dimensions, as a background field becomes concentrated on
the surface on which the Dirichlet boundary condition would eventually
hold, the Casimir energy diverges.  This result implies that the
energy depends in detail on the properties of the material, which are
not captured by the idealized boundary conditions.  This divergence
does not affect the force between rigid bodies, but it does invalidate
calculations of Casimir stresses based on idealized boundary
conditions.}

\leftline{\it \small Keywords:~\parbox[t]{15cm}{
Energy densities, Green's functions, renormalization, Casimir effect}}

\leftline{\it \small PACS:~\parbox[t]{15cm}{ 11.10.-z, 11.10.Gh, 
03.70.+k}}

\newpage

\bigskip
\section{Introduction} 

The development of powerful experimental methods to measure the
Casimir force and its potential importance in micromechanical
devices have stimulated new interest in the physical principles
underlying the Casimir effect \cite{expt1,expt2,expt3,bmm}.  The
Casimir force was originally discovered by Casimir and Polder in
their study of the long range, relativistic tail of the van der Waals
force between neutral atoms \cite{CasimirPolder}.  Immediately
afterwards, Casimir showed that the force between grounded,
conducting plates could be understood as a modification of the
quantum fluctuations of the electromagnetic fields forced to obey
conducting boundary conditions on the plates \cite{casimir}.

The intriguing idea of a force generated by the modification of the
zero point energy of a fluctuating quantum field due to the imposition
of a boundary condition has generated continuing theoretical interest
in generalized ``Casimir problems'' \cite{MT}. The general idea is to
specify a) an otherwise free quantum field (scalar, Dirac, gauge, {\it
etc.\/}), b) a boundary condition (Dirichlet, confining, conducting,
{\it etc.\/}), and c) a geometry (parallel plates, spherical,
cylindrical, {\it etc.\/}), and to calculate the forces, pressures,
and other potential observables that arise when the field is forced to
obey the boundary condition on the surface.

These Casimir problems are idealizations in which the physical
interactions between the fluctuating fields and matter have been
replaced {\it ab initio\/} by boundary conditions.  A real material
cannot constrain modes of the field with wavelengths much smaller than
the typical length scale of its interactions.  The interactions become
negligible at wavelengths less than certain physically determined
cutoffs.  In contrast, a boundary condition constrains all modes.  To
calculate the Casimir energy it is necessary to sum over the zero
point energy of all modes.  This sum is highly divergent in the
ultraviolet and these divergences depend on the boundary conditions. 
Subtraction of the vacuum energy in the absence of boundaries removes
only the worst divergence (quartic in three space dimensions).  We ask
whether the Casimir energy and other potential observables can be
defined independent of the cutoffs that characterize the actual
interactions between the fluctuating fields and the matter.  If so,
then one can define an abstract Casimir problem, depending on the
field, the boundary condition, and the geometry alone.  If not, then
the Casimir energy is unavoidably entangled with the detailed material
physics at hand.  We study the simple case of a scalar field, $\phi$,
the Dirichlet condition, $\phi=0$, and three geometries: single plate,
parallel plates and sphere.  We believe this case illustrates general
principles that affect all Casimir calculations.

We find that the Casimir energy always depends on the cutoffs and
diverges as the cutoffs go to infinity.  The divergences in three
dimensions are even more severe than the divergences in one and two
dimensions described in earlier work \cite{Graham:2002fw}.  We show
that these divergences can be isolated in low-order Feynman diagrams,
which can be studied with standard methods of renormalized quantum
field theory.  It is important to emphasize, however, that the Casimir
force between rigid bodies, which is the only thing that has been
measured experimentally, is cutoff independent and therefore finite in
the limit of infinite cutoff.  The most important example of an
observable that is unavoidably cutoff dependent is the ``Casimir
pressure''  ($-\partial E/\partial A$), which we study
for the case of a sphere.

The method of renormalization in continuum quantum field theory
\emph{without boundaries} (QFT) provides the only physical way to
regulate, discuss, and eventually remove divergences.  We propose to
replace a boundary condition by a renormalizable coupling between the
fluctuating field and a non-dynamical background field representing
the material.  When the background is smooth and the coupling is
finite, then renormalization ensures that the zero point energy of the
fluctuating field relative to the vacuum is finite.  In the limits
that the background becomes sharply peaked on a surface and the
coupling becomes strong, all modes of the fluctuating field will obey
a boundary condition on the surface.  If a physical observable like
the pressure diverges in this limit then the observable depends on the
cutoffs that in reality keep the background from becoming arbitrarily
sharp and the coupling from becoming arbitrarily strong.  Our
prescription is: a) fix the background, b) compute the renormalized
zero point energy, c) take the ``boundary condition limit''.  This is
to be contrasted with the standard approach: a$^{\prime}$) take the
boundary condition limit, b$^{\prime}$) compute the zero point energy. 
The two approaches disagree for all surfaces (plates and spheres) and
in all dimensions that we have studied.  We believe that our
prescription for the order of limits is the physical one, and that the
result obtained by imposing the boundary condition {\it ab initio\/}
should be discarded when it disagrees with it.
 
Here we study the imposition of Dirichlet boundary conditions on a
scalar field.  It is straightforward to write down a QFT describing a
renormalizable interaction between the fluctuating scalar field $\phi$
and a static, non-dynamical background field $\sigma(\vek x)$,
\begin{equation}
\mathcal{L}_{\rm int}(\phi,\sigma)=-\frac{1}{2}\lambda\,
\sigma(\vek{x})\,\phi^{2}(\vek{x},t)
\label{lint}
\end{equation}
and to choose a limit involving the shape of $\sigma(\vek x)$ and the
coupling strength, $\lambda$, between $\phi$ and $\sigma$ that
produces the desired boundary conditions on specified surfaces.  The
 interaction strength  is characterized by two parameters: the
width $\Delta$ of the background  and the strength $\lambda$. 
In the limit $\Delta\to 0$ the background becomes a surface
$\delta$-function.  We refer to this as the ``sharp limit.''  Once the
sharp limit has been taken, the limit $\lambda\to\infty$ enforces the
Dirichlet boundary condition on all modes of the fluctuating field. 
We refer to this as the ``strong'' limit.  The boundary condition
emerges on the surface only in the limit that the background becomes
both sharp and strong.  Both $\Delta$ and $\lambda$ represent physical
cutoffs characteristic of the material with which the fluctuating
field interacts.  We identify $\Delta$ with the physical thickness of
the surface and $\lambda$ plays a role similar to the plasma
frequency: modes with frequency much larger than the scale determined
by $\lambda$ are not constrained at the boundary.

Ideally, we seek a Casimir energy that reflects only the effects of
the boundary conditions and not any other features of $\sigma(\vek
x)$.  Therefore we do not specify any action for $\sigma$ except for
the standard counterterms induced by the $\phi$-$\sigma$ interaction. 
The coefficients of the counterterms are fixed by renormalization
conditions applied to perturbative Green's functions.  Having been
fixed in perturbation theory, the counterterms are fixed once and for
all and must serve to remove the divergences that arise for any
physically sensible $\sigma(\vek x)$.  Moreover, the renormalization
conditions are independent of the particular choice of background
$\sigma(\vek x)$, so it makes sense to compare results for different
choices of $\sigma(\vek x)$, {\it i.e.\/} different geometries.  In
the simple theory we are considering, the only counterterms
are ${\cal L}_{\rm ct\,1}(\sigma)=c_{1}\sigma(\vek x)$, which
renormalizes the tadpole graph and, in three dimensions, ${\cal
L}_{\rm ct\,2}(\sigma)=c_{2}\sigma^{2}(\vek x)$, which renormalizes
the self-energy.   

\begin{figure}
\vskip1.0cm \centerline{
\BoxedEPSF{ 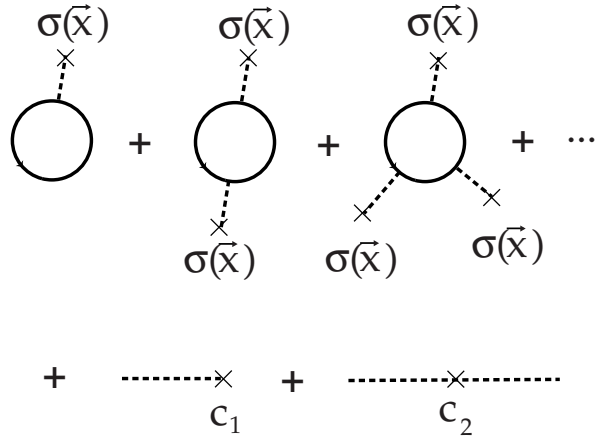 scaled 650}}
\caption{\label{loops}\sl The Feynman graph series for the Casimir 
energy.}
\end{figure}

The vacuum polarization energy --- the Casimir energy, $E[\sigma]$ ---
for the field $\phi$ in the background $\sigma$ is given by the sum of
all one-loop Feynman diagrams plus the contributions of the
counterterms, as shown in Fig.~\ref{loops}.   

The diagrams generate a representation for
$E[\sigma]$ as a power series in $\lambda$.  It is easy to show
that
\begin{itemize}
	\item for piecewise continuous $\sigma$ (characterized by a width
	$\Delta$), the contribution to the renormalized Casimir energy at
	each order in $\lambda$ is finite, and 
	\item the sum over the terms in the power series in $\lambda$ gives a
	finite, renormalized Casimir energy for any fixed $\lambda$ and
	$\Delta$.
\end{itemize}
The first result is a general consequence of the renormalizability of 
the $\phi$-$\sigma$ field theory.  The second result can be 
demonstrated explicitly by rewriting the Casimir energy as a 
sum/integral over scattering data (bound state energies and scattering 
phase shifts) in the fashion of Refs.~\cite{us,GJW,dens}, where it is 
manifestly finite.
 
The thickness $\Delta$ sets the scale for the Fourier components,
$\tilde\sigma(\vek p)=\int
d^{\,D}x\,\sigma(\vek{x})\exp(i\vek{p}\cdot\vek{x})$ in D space
dimensions, of the background field , which are integrated in the
one-loop diagrams of Fig.~\ref{loops}.  As $\Delta\to 0$ the Fourier
integrals in low-order diagrams diverge.  This simple mechanism is
responsible for the cutoff dependence of the Casimir energy.  Because
it occurs in low-order diagrams ($\mbox{order}\le 3$ for three
dimensions or less) we can examine the divergences with well-known and
universally accepted methods of analysis\cite{PS}.  This analysis
forms the technical core of this paper.

The simplest diagram, the tadpole, is momentum independent.  In our
renormalization scheme it is canceled exactly by the counterterm
${\cal L}_{\rm ct\,1}$ leaving no contribution to the Casimir energy
at order $\lambda$.  The two-point function has a simple
representation in terms of $\tilde\sigma(\vek p)$.  In two dimensions,
we have
\begin{equation}
E_{\rm 2D}^{(2)}[\si] =-\frac{\lambda^2}{16\pi}
\int \frac{d^{\,2}p}{(2\pi)^2} 
\tilde{\sigma}(\vek p)\tilde{\si}(-\vek p)\,
\frac{1}{|\vek p|}\tan^{-1}\frac{|\vek p|}{2m}\, ,
\label{2Dtwopoint}
\end{equation}
and in three dimensions, after renormalization we have
\begin{equation}
E_{\rm 3D}^{(2)}[\si] = \frac{\lambda^2}{64\pi^2} \int
\frac{d^{\,3}p}{(2\pi)^3} \tilde{\sigma}(\vek p)
\,\tilde{\sigma}(-\vek p) \int_0^1 dx \
{\ln}\frac{m^2+x(1-x)\vek{p}^2}{m^2+x(1-x)\mu^2}\, .
\label{3Dtwopoint}
\end{equation}
Note the dependence of $E_{\rm 3D}^{(2)}[\si] $ on the (spacelike)
renormalization scale $\mu$.  In the first part of Section 3 we show
that these contributions to the Casimir energy diverge like $\ln
\Delta$ and $(\ln \Delta)/\Delta$ in the sharp limit.  There is no
mechanism available to cancel these divergences.  In two dimensions,
if we use the counterterm ${\cal L}_{\rm ct\, 2}(\sigma)$ with finite
coefficient $c_{2}$ to make a finite renormalization of the two-point
function, the divergence only gets worse, as discussed in Section 3.2. 
In three dimensions we have already renormalized the theory as the
$\mu$ dependence of $E_{\rm 3D}^{(2)}[\si] $ attests and yet the
divergence in the sharp limit persists.  We show that these cutoff
dependences affect the Casimir pressure.

The reader may still worry that some more subtle interpretation of
renormalization will still succeed in removing the divergence,
especially in $E_{\rm 3D}^{(2)}[\si]$ which had a primitive (loop)
divergence prior to renormalization even for fixed $\Delta$.  To
address that concern we study the three-point function in three
dimensions, for which there is no loop divergence.  We show that this
diagram also diverges (like $\ln \Delta$) in the sharp limit --- the
situation is analogous to the two-point function in two dimensions. 
There is no mechanism to remove this divergence at any stage in the
formulation of the theory, and it contributes to the Casimir pressure
as well.  Thus the Casimir pressure on a sphere is unavoidably cutoff
dependent both in two and three dimensions.

We have made many consistency checks on this calculation.  In addition
to the analytical  analysis of the $\Delta\to 0$ limit given in
Sections 3 and 4, we have computed (in Section 5) the diagrams
\emph{in the sharp limit} with a $\delta$-function source but with a
frequency cutoff on $\tilde\sigma(\vek p)$.  This is a simpler
computation and it yields the same result.  Finally we have used the
``interface'' formalism developed in Ref.~\cite{GJQW} (in the
Appendix) to generalize from one to two to three dimensions, again
confirming the result of the direct analytical analysis.
 
 The fact that the energy of a fluctuating field diverges when
a boundary condition is imposed has been known for many years
\cite{C&D, Candelas}.  However divergences are common in quantum field
theory and renormalization was invented to remove them.  Special
calculational methods have been developed to eliminate the extra
divergences that arise in the presence of boundaries, which give
finite, cutoff-independent Casimir energies\cite{MT}.  These methods
do not address the crucial question of whether the discarded
divergences reflect physically significant cutoff dependence or are
merely unphysical artifacts of the calculational method.  Several
 studies of these divergences have been made.  Symanzik studied
the divergences of a scalar field theory in the presence of a surface
interaction that imposes the boundary condition $\phi=0$
\cite{Symanzik}.  He showed that it is possible to introduce a set of
surface counterterms sufficient to cancel all the divergences that
arise in this theory.  There are many independent counterterms and
their forms are complex.  However, the counterterms available in a
renormalizable quantum field theory are few in number and simple in
form, merely polynomials in the field and its derivatives. 
There is no freedom to introduce Symanzik's {\it ad hoc\/} surface
counterterms if one views the boundary condition as arising from an
underlying renormalizable quantum field theory (as it must). 
Therefore his methods cannot be used to explore the physical
implications of the divergences generated by boundary conditions.

 The work of Candelas\cite{Candelas} and Candelas and
Deutsch\cite{C&D} is similar to our both in spirit and conclusions. 
In Ref.~\cite{C&D} they studied the Casimir energy and energy density
for scalar and electromagnetic fields in the presence of boundaries. 
Candelas later refined this work and extended it to boundaries between
dielectrics.  They point out that the vacuum polarization energy
density generally diverges as one approaches a boundary and state that
these divergences are not the ones that can be cancelled by standard
renormalization methods.  They note that the famous special cases
where the divergences cancel -- the conformal scalar field near a
planar boundary\cite{Brown} and the electromagnetic field near a plane
or a sphere\cite{Boyer} -- are indeed exceptional.  The cancellations
fail and the Casimir energy density is divergent if the surface has
(arbitrarily small) shape imperfections.  Candelas and Deutsch
speculate that divergences are generic and that a more realistic
treatment of the boundary would confirm this.  In Ref.~\cite{Candelas}
Candelas develops a model of a dielectric in order to examine the
divergences in Casimir energies under more realistic conditions.  He
finds that divergences are generic.  The divergences that arise in the
dielectric case are worse than those found when conducting boundary
conditions are applied {\it ab initio\/}.  Candelas warns that his
results cast doubt on Casimir calculations which begin with idealized
boundary conditions.  However, in the years following Candelas's work
this warning seems largely to have been ignored.  The dielectric model
used by Candelas does not correspond to a renormalizable continuum
quantum field theory.  \footnote{  Candelas models the
dielectric in terms of elementary electric and magnetic oscillators,
which could in principle be embedded in a renormalizable theory. 
However the \emph{effective field theory} obtained by integrating out
the oscillators to obtain a dielectic is no longer renormalizable. }
Therefore one cannot state for certain which divergences can or cannot
be cancelled by counterterms.  Also, when he studies conductors or the
Dirichlet problem, Candelas imposes boundary conditions {\it ab
initio\/}, introducing the possibility of missing physically
significant divergences.  Our work can be regarded as a reformulation
of Candelas's critique of Casimir calculations in the context of
renormalizable quantum field theory.  In this context there is no
longer any reason to doubt the physical significance of the cutoff
dependence we have found. 
 
 Finally  our results disagree with  the recent results
of Ref.~\cite{Milton}, which also considers low-order Feynman
diagrams, but obtains a finite contribution to the stress in the
boundary condition limit.  The specific disagreements are discussed in
the Conclusions after we have introduced the conceptual framework
necessary for the discussion.
 
We note that some of these results have appeared in a brief
introduction to our work \cite{Graham:2002fw}  and in a
conference presentation \cite{Jaffe:2003ji}, where the one-dimensional
case was discussed in complete detail.  A summary of the two
dimensional case was also given in \cite{Graham:2002fw}.  The
divergences of the Casimir energy in one dimension are less severe,
but still prevent one from defining a Casimir energy in the Dirichlet
boundary condition limit.  Finally, we note that subtleties similar to
the ones we have considered were addressed in the context of
dispersive media in Ref.  \cite{Barton}

\section{General structure of the Casimir energy}

In this section we outline the calculation of the Casimir energy of a
scalar field in a piecewise continuous scalar background field in
two or three space dimensions ($D=2,3$).  We review the Feynman diagram
expansion and renormalization issues and summarize the scattering data
approach which sums all Feynman diagrams and gives an unambiguous
renormalized vacuum polarization energy.  We take a first look at how
divergences can arise in the renormalized Casimir energy as the
background becomes singular and identify the Feynman diagrams that
play a central role.

We consider the theory of a real scalar field $\phi$ coupled to a
time-independent background field $\sigma(\vek x)$, described by the
Lagrangian,
\begin{equation}
    \mathcal{L} = \frac{1}{2} \partial_\mu\phi\,\partial^\mu\phi -
    \frac{m^2}{2} \phi^2 - \frac{\lambda}{2} \phi^2 \sigma(\vek{x}) +
    \mathcal{L}_{\rm CT}[\sigma]\,.
    \label{lag}
\end{equation}
${\cal L}_{\rm CT}[\sigma]$ is the counterterm Lagrangian required by
renormalization.

The Casimir energy is defined as the vacuum energy in the presence of
$\sigma$ minus the vacuum energy in its absence.  It can be written
formally as the sum over the shifts in the zero-point energies of all
the modes of $\phi$ relative to the trivial background $\sigma=0$, $
E_{\rm bare}[\sigma]= \frac{1} {2}\sum_n (\omega_n[\sigma] -
\omega_n^{(0)})$.  Equivalently, using the effective action formalism,
$E_{\rm bare}[\sigma]$ is given by the sum of all  one-loop Feynman
diagrams with at least one external $\sigma$ field, as shown in
Fig.~\ref{loops}. The low-order diagrams generate divergences,
which are canceled by the counterterms in ${\cal L}_{\rm CT}$. 
Combining its contribution to the energy, $E_{\rm CT}[\si]$, with
$E_{\rm bare}[\sigma]$ yields the renormalized Casimir energy
$E[\sigma]$. 

We have introduced $\sigma(\vek x)$ as a non-dynamical, external
field, solely to constrain $\phi$.  In addition to the $\phi$-$\sigma$
coupling we include only the counterterms induced by the
renormalization process.  Of course one could always add non-trivial
dynamics for $\sigma$ in the form ${\cal L}[\sigma] =
\half(\p\si)^{2}-V(\si)$.  Such terms are undoubtedly present in a
real material, where they correspond to the self-interactions of the
material.  However, since we want to know whether the vacuum energy
associated with the fluctuations of $\phi$ can be isolated from the
rest of the problem, we suppress any additional dynamics associated
with $\si$.
 
Because we are calculating in a renormalizable quantum field theory,
$E[\sigma]$ will be finite for any smooth $\sigma$ and finite
$\lambda$.  Renormalization is straightforward.  First the theory is
regulated.  Next, superficially divergent Feynman diagrams are
identified.  Only the one-point function (the ``tadpole'') and the
two-point function (the $\si$ self-energy) are superficially divergent
for $D\le 3$.  The divergences are local and can therefore be canceled
by counterterms of the form
\begin{equation}
    {\cal L}_{\rm CT}[\si]= c_{1}(\eta)\sigma +
    c_{2}(\eta)\sigma^{2}
    \lab{ct}
\end{equation}
Here $\eta$ is a regulator, for example the deviation from the
physical spacetime dimension in dimensional regularization.  The
constants $c_{1}$ and $c_{2}$ are fixed by renormalization conditions. 
We choose the ``no tadpole'' condition $\langle\si\rangle=0$ to fix
$c_{1}$, and we require that the renormalized two-point function,
$\bar\Gamma_{2}$, vanish at $p^{2}=-\mu^{2}$ to fix $c_{2}$.  Of
course any other choice of renormalization scheme can be related to
ours by appropriate renormalization group transformations.  We wish to
stress that the renormalization process fixes the counterterms
completely, independent of $\si$.  So, for example, the counterterm
$c_{1}$ is given by,
\begin{equation}
    c_{1}=\lambda \frac{m^{n-2}}{(4\pi)^{\frac{n}{2}}}\Gamma\left(
    \frac{2-n}{2}\right)
\end{equation}
in $n$ spacetime dimensions.   No {\it ad hoc} modifications of the
counterterms can be introduced to cancel an unanticipated divergence.

Although the Casimir energy can be formally represented as the sum of
all one-loop Feynman diagrams, only the first few can be computed
directly.  As the number of external lines increases so does the
number of integrations over the external momenta carried by
$\tilde\si(\vek p)$, the Fourier transform of $\sigma(\vek x)$.  In
practice it becomes tedious to compute beyond the three-point function
directly.  Furthermore the power series in $\lambda$ generated by the
Feynman diagrams may not converge.  We rely instead on methods that
allows us to compute the Casimir energy of the background
configuration \emph{exactly} for any $\lambda$, while still performing
all the necessary renormalization in the perturbative sector.  The
full Casimir energy can be computed exactly using phase shift methods,
\cite{us,GJW,dens}, by summing the derivative expansion
\cite{Chan,Dunne}, or through analysis of Green's functions
\cite{dens,Baacke}.  All these methods show that once the counterterms
corresponding to low-order diagrams have been included, the total
Casimir energy $E[\si]$ for a smooth background is finite.  Expanding
this result in powers of $\lambda$, we recover the diagrammatic
expansion and can isolate the effects of the counterterms.

Before analyzing the contributions of order $\la^{2}$ and $\la^{3}$ in
detail, we give an overview of the possible divergences in $E[\si]$
when the background $\si$ becomes sharp and strong.  We will consider
a square barrier of width $\De$ and height $1/\De$.  Any smooth shape
that approaches a surface delta function as $\Delta\to 0$ would do.
\begin{itemize}
	\item The tadpole diagram is local, {\it i.e.\/} proportional to
	$\si(\vek x)$, and therefore is canceled entirely by the
	counterterm ${\cal L}_{1}=c_{1}\si(\vek x)$.  The resulting rule
	for calculation is merely to drop the contribution of first order
	in $\la$ to the Casimir energy.  No matter how singular the
	source, the contribution of order $\la$ to $E$ vanishes as a
	result of our choice of renormalization scheme.
    
	\item The contribution of order $\la^{2}$ comes entirely from the
	renormalized two-point function, which can be computed using
	standard methods.  So the rule for calculation is to remove the
	terms of order $\la^{2}$ from the scattering data and replace them
	by $E^{(2)}[\si]$.  The expressions for $E^{(2)}[\si]$ in two and
	three space dimensions have been given in eqs.~(\ref{2Dtwopoint})
	and eq.~(\ref{3Dtwopoint}) respectively.
    
	\smallskip\noindent $E^{(2)}[\si]$ can diverge if the background
	$\si$ has a Fourier spectrum that vanishes too slowly at large
	$p$.  This is precisely what happens when $\De\to 0$.  Thus,
	examination of the renormalized two-point function alone is enough
	to establish that the renormalized Casimir energy diverges in the
	Dirichlet limit.
    
	\item The higher order (in $\la$) contributions to $E[\si]$ are
	guaranteed to be finite for smooth enough $\si$ and finite $\la$
	because only the tadpole and self-energy diagrams have primitive
	divergences in the underlying field theory for $D\le 3$.  We find
	that $\si(\vek x)$ which are piecewise continuous are smooth
	enough to give a finite renormalized Casimir energy.  Thus the
	renormalized Casimir energy is finite for any fixed $\De$ and
	$\la$.

	\smallskip\noindent In two dimensions there are no divergences
	beyond the two-point function even in the sharp limit.  However,
	in three dimensions as the background field becomes sharp, {\it
	i.e.\/} as $\De\to 0$, the three-point function diverges as well. 
	In Section 4 we calculate the three-point function in the
	background of parallel plates of width $\De$ and also in the
	background of a spherical shell of with $\De$.  We study the limit
	$\De\to 0$ and show that this contribution to $E$ diverges.
	
	In Section 5, we calculate the contribution to the vacuum
	polarization energy at 
	$n^{\rm th}$ order in $\lambda$ for a surface $\delta$-function in
	three dimensions for both plane and spherical
	geometries.  We find
	a result that is finite for $n>3$, but diverges
	logarithmically for $n=3$, a divergence that cannot have anything
	to do with choice of renormalization scheme or other subtleties. 

\end{itemize}

To summarize: Potential divergences in the Casimir energy in the
Dirichlet limit are isolated in the two-point function in two
dimensions and in the two and three-point functions in three
dimensions.  These contributions are analyzed in depth in the next two
sections for single plate, parallel plate and spherical geometries in
two and three dimensions.

\section{The Two-Point Function}
\bigskip
 
In this section and the next we analyze the low-order Feynman diagrams
that are potentially divergent in the sharp limit, $\Delta\to 0$.  We
treat the case of a square barrier, although any one-parameter family
of functions which approaches a $\delta$-function as $\De\to 0$ would
do.  In this section we study the renormalized two-point function as
defined in eqs.~(\ref{2Dtwopoint}) and (\ref{3Dtwopoint}).  While this
analysis alone is sufficient to establish a divergence in the total
energy, the fact that  in three space dimensions it requires
renormalization may make some readers uneasy.  The three-point
function, while more difficult to compute, has no such complication
and we will study its divergences in the next section.

We study three geometries:  
a single plate 
\begin{equation}
    \sigma_{\vert}(z)=\frac{1
    }{\De}\Big(\theta(z+\Delta/2)-\theta(z-\Delta/2)\Big)\, ,
    \label{sig1}
\end{equation}
parallel plates  separated by a distance $2L$ (both configurations 
in D($\le 3$) space dimensions), 
\begin{equation}
    \sigma_{\parallel}(z)=\frac{1}{\De} \Big(\theta (|z|-L+\De/2 ) 
        -\theta (|z|-L-\De/2) \Big)\, ,
    \label{sig2}
\end{equation}
and a  sphere  of radius $R$ in D=3 space dimensions,
\begin{equation}
    \sigma_{\circ}(r)=
         \frac{3}{4 \pi  \left((R+\De)^{3}-R^{3}\right)} 
        \Big(\theta(r-R)-\theta(r-R-\De)\Big)\, .
    \lab{sig3}
\end{equation} 
We have chosen to normalize $\si$ analogously for plates and spheres. 
For plates we have normalized $\si$ so its integral across each plate
is proportional to the surface area and independent of the width $\De$
and the number of dimensions, {\it i.e.\/} we require $\int
dz\si(z)=N$, where $N=1\ \mbox{or} \ 2$, is the number of plates
present.  For $\si_{\vert}$ and $\si_{\parallel}$ we study ${\cal E}$,
the Casimir energy per unit area in three dimensions or per unit
length in two dimensions.  In the case of a spherical shell of radius
$R$ we have normalized $\si$ so that
\begin{equation}
	\int d^{\, 3}\!x\si_{\circ}(\vek x)=1
	\lab{norm1}
\end{equation}
independent of the width $\De$.   We are aware of the fact that 
 different normalization choices imply different dimensions for
$\lambda$.  These can be easily introduced, for example, by introducing
appropriate powers of $m$, the mass of the fluctuating particle. 

The choice of normalization has little significance for single or
parallel plates.  However it does have physical consequences when we
measure the stress on a circular or spherical shell.  To measure the
stress, it is necessary to compare the energies of shells of radius
$R$ and $R+\delta R$.  To make this comparison, one must decide how
the matter behaves as $R$ changes.  Unfortunately, one of the
shortcomings of the traditional boundary condition approach is that it
tells us nothing about the physical properties of the material.  So we
must consider a range of possible behaviors.  With the normalization
chosen above, $\int dr \si_{\circ}(r)\sim 1/R^{2}$ in three
dimensions.  This normalization corresponds to keeping the volume of
$\si$ fixed as the shell is expanded, as if it were a physical
substance.  Another possibility would be to hold $\int
dr\si_{\circ}(r)$ fixed as $R$ varies.  We consider both possibilities
when discussing the stress on a shell.  We will see that it is not
possible to choose any normalization of $\sigma$ that would eliminate
divergences in the renormalized Casimir stress on a sphere in three
dimensions.

It is convenient to work in momentum space when computing n-point
functions.  The background fields in momentum space are given by the
Fourier transforms of eqs.~(\ref{sig1}), (\ref{sig2}), (\ref{sig3}) in
the case of the single plate, parallel plates and sphere respectively:
\begin{eqnarray}
	\tilde\sigma_\vert (\vek q)&=& \Pi_{i=1}^{D-1} 
        \left[ (2\pi)\delta(q^i)\right] \frac{2}{q_z\De}\,
         {\rm sin}\left(\frac{q_z\De}{2}\right)\, , 
	\label{ffoneplate} \\
	\tilde\sigma_\parallel (\vek q)&=& \Pi_{i=1}^{D-1} 
        \left[ (2\pi)\delta(q^i)\right]\frac{4}{q_z\De}\,\cos(q_zL)\,
	{\rm sin}\left(\frac{q_z\De}{2}\right)\, , 
	\label{ffplate} \\
	\tilde\sigma_\circ (\vek q)&=& \frac{3}{ q^3}
	\frac{qR\left[{\cos}(qR) -(1+\De/R){\cos}(q(R+\De))\right]
	-{\sin}(qR)+{\sin}(q(R+\De))} {(R+\De)^3-R^{3}} \, ,
	\label{ffsphere}
\end{eqnarray}
where $q=|\vek q|$.

Most of our interest lies in three dimensions where there has been
some controversy about the nature of the divergences in this
problem \cite{Milton}.  At the end of this section we present analogous
results in two dimensions for comparison.  The two-point function
contribution to the Casimir energy can be computed analytically for a
\emph{massless} scalar field.  Therefore we begin with the massless
case.

\subsection{Massless Case}

In two dimensions the two-point function contribution to the Casimir
energy is infrared divergent in the massless case.  So in this
subsection we confine ourselves to three dimensions.   
In three dimensions the renormalized two-point function was given in 
eq.~(\ref{3Dtwopoint}), and simplifies in the massless
case to 
    \begin{equation}
	E_{\rm 3D}^{(2)}[\si] =\frac{\lambda^2}{64\pi^2} \int
	\frac{d^{\,3}p}{(2 \pi)^3}
	\tilde{\sigma}(\vek p) \,\tilde{\sigma}(-\vek p) 
	{\ln}\frac{\vek{p}^2}{\mu^2}\, ,
    \end{equation} 
 
For the single plate, eq.~(\ref{sig1}), we obtain
\begin{eqnarray}
	{\cal E}^{(2)}_{\rm 3D}(\si_{\vert}) &=&
	\frac{\lambda^2 }{16\pi^2\De^2} 
	\int_{-\infty}^{\infty} \frac{dp}{2 \pi}\, \frac{1}{p^2} \sin^2{\left(p
	\De/2\right)} \ln{\left(p^2/ \mu^2 \right)} \cr &=& -
	\frac{\lambda^2 }{32\pi^2 \De} \left(\ln{(\mu \De)} + \gamma - 1
	\right)\,,
\label{e2single}
\end{eqnarray}
for the energy per unit area, where $\gamma = 0.577 \ldots$ is Euler's
constant.  From this we conclude that the Casimir energy per unit area
of a single plate diverges as $\frac{1}{\De}\ln\De$ in the \emph{sharp
limit}, $\De \to 0$.   This is the first example of a general
result: if we normalize the background field $\sigma$ to the surface
area of the body under consideration, then the form of the leading and
the first subleading divergence is always
\begin{equation}
	E^{(2)}_{\rm 3D}= -\frac{\lambda^{2}}{32\pi^{2}}
	\frac{A}{\De}(\ln(\mu\De) +\gamma-1),
\end{equation}
where  $A$ is the area.  In the cases
of single and parallel plates there are no further divergences in the
sharp limit, but in the case of a sphere there is a further sub-subleading
divergence that is not universal. 

Next  consider the case of parallel plates, eq.~(\ref{sig2}),
\begin{equation} 
	{\cal E}^{(2)}_{\rm  3D}(\si_{\parallel}) = 
	\frac{\lambda^2}{8\pi^2\De^2}  
        \int_{-\infty}^{\infty} \frac{dp}{2 \pi}\,
	 \frac{1}{p^2} \left( 1+\cos{(2pL)}\right) 
	   \sin^2{\left(p\De/2 \right)} 
	 \ln{\left(p^2/\mu^2 \right)} 
	\label{e2plates1}
\end{equation}
This integral can be carried out by
deforming the integration contour in the complex plane,
\begin{eqnarray}
	{\cal E}^{(2)}_{\rm  3D}(\si_{\parallel}) &=&
        -\frac{\lambda^2}{16\pi^2 \De} \left\{\ln{(\De
	  \mu)} + \gamma-1 
	+ \left(\half+\frac{L}{\De}\right)\right.
	 \ln(1 + \De /2L) \cr && \hspace{3cm} \left.
        -  \left(\half-\frac{L}{\De}\right)
	 \ln(1 - \De /2L)\right\}\, ,  
\label{e2plates2}
\end{eqnarray}
 It is easy to check that this result has the proper behavior as the
 plates separate to infinity or coalesce.  As $L\to\infty$, the energy
 per unit area approaches twice the result for a single plate of unit
 strength (see eq.~(\ref{e2single})).  And as the plates coalesce at
 $L=\De/2$, ${\cal E}^{(2)}_{\rm 3D}(\si_{\parallel})$ becomes equal
 to the energy per unit area of a single plate of strength two with
 width $2\De$, as expected.  As expected, the divergence in ${\cal
 E}^{(2)}_{\rm 3D}(\si_{\parallel})$ is twice that in ${\cal
 E}^{(2)}_{\rm 3D}(\si_{\vert})$.

 Although the Casimir energy for two parallel plates diverges as
$\Delta\to 0$, the divergent terms are independent of $L$, explicitly,
\begin{equation}
	{\cal E}^{(2)}_{\rm  3D}(\si_{\parallel}) =
        -\frac{\lambda^2}{16\pi^2 \De} \left\{\ln{(\De
	  \mu)} + \gamma-1 +\frac{\Delta}{4L} + {\cal O}\left(
	  \frac{\Delta^{2}}{L^{2}}\right)\right\}\, ,
\label{e2limit}
\end{equation}
and therefore the two-point contribution to the Casimir \emph{force}
remains finite in the sharp limit.  Also, only the
$L$-\emph{independent} terms depend on the renormalization scale
$\mu$.  That is, this force does not have renormalization ambiguities.

Next, we turn to the spherical shell.  Using eq.~(\ref{sig3}), we 
obtain,
 \begin{eqnarray}
	   E^{(2)}_{\rm  3D}(\si_{\circ}) &=&  
	   -\frac{\lambda^2}{256 \pi^3 R
	   \De^2(1+\eps+\frac{1}{2}\eps^{2})^2}
	   \left\{ \frac{2}{3}  \ln{\frac{4(1+\eps)}{(2+\eps)^2}} -
	    2 \eps  \ln{\frac{2+2 \eps}{2+\eps}} \right. 
	   \nonumber \\ && \left.  + 2  \eps(\ln{(\mu \De)} + \gamma -
	   1) +  \eps^2 \left(- \frac{4}{3} + 2 \ln{\frac{2+2
	   \eps}{2+\eps}} + 2 (\ln{(\mu \De)} + \gamma -
	   1) \right) \right.  \nonumber \\
	   && \left.  + \eps^3\left( -\frac{8}{9} + \frac{2}{3}
	   \ln{\frac{2+2\eps}{2+\eps}} + \frac{2}{3} 
	   (\ln{(\mu \De)} + \gamma -1) \right) \right\}.
\end{eqnarray}
where $\eps=\De/R$.    The terms that diverge in the
limit $\De \to 0$ can be singled out
\begin{eqnarray}
	   E^{(2)}_{\rm  3D}(\si_{\circ}) &=& 
           -\frac{\lambda^2}{128 \pi^3 R^2} \left(\frac{1}{\De} \left(\ln{(\mu
	   \De)}+ \gamma-1 \right) - \frac{\ln{(\mu
	   \De)}}{R} \right) +
	   \textrm{~terms~finite~for~} \De \to 0\,.\,\,
	   \label{div3}
\end{eqnarray}
If we correct for the different normalizations, we find that the
leading and the first subleading divergent contribution is the same
(per unit area) as for the plate(s).  However its physical
consequences are quite different.  In the case of an isolated sphere
the stress,
\begin{equation}
	\mathcal{P}^{(2)}_{\rm  3D}(\si_{\circ})=-
        \frac{\p E^{(2)}_{\rm  3D}(\si_{\circ})}{\p R}
\end{equation}
is the quantity of interest.  Apparently the stress diverges as
$\frac{1}{R^{3}\De}\ln \De$ as $\De\to 0$.  Had we normalized
$\si_{\circ}$ in the same way as $\si_{|}$ and $\si_{\parallel}$, so
that $\int dr\si_{\circ}(r)\sim$ constant, the $R$ dependence of the
prefactor in eq.~(\ref{div3}) would have changed from $\propto
1/R^{2}$ to $\propto R^{2}$ .  The stress changes sign, however it
would still diverge as $\Delta\to 0$.  This highlights the fact that
the stress depends in detail on the surface dynamics.  

 The reader might be tempted to eliminate the $R$ dependence
from the divergences in eq.  (\ref{div3}) by choosing a different
normalization for the background $\sigma$ and redefining $R\to
R+\De/2$.  Such a transformation is possible, and would render the 
Casimir pressure from the two-point function finite.   However, as we 
shall see below, the three-point function also contributes a divergent 
Casimir pressure, and no transformations can eliminate all of the 
divergences. 

\subsection{Case of Non-zero Mass} 

\subsubsection{Three Dimensions}
It has been claimed that the divergences in the sharp limit are an
artifact of the specialization to massless bosons
\cite{Milton}.  It is certainly true that low-order
Feynman diagrams have infrared divergences in low dimension.  However
these are not the divergences that plague the Casimir energy.  To show
that the divergences of the previous section persist when $m\ne 0$ we
present the case of a single plate.  The other geometries of interest
behave analogously.

For a scalar boson of mass $m$ in three dimensions we have to evaluate 
\begin{eqnarray}
	{\cal E}^{(2)}_{\rm  3D}(\si_{\vert})&=& 
        \frac{\lambda^2}{16\pi^2\De^2} \int_{-\infty}^{\infty} 
        \frac{dp}{2 \pi}\, \frac{1}{p^2}
	\sin^2{\left(p \De/2\right)} \int_0^1 dx\, \ln{\left(\frac{m^2+
	x(1-x)p^2}{m^2+x(1-x)\mu^2} \right)} \nonumber \\
	&=& - 	\frac{\lambda^2}{32\pi^2\De} \Bigg\{
	\ln{\left(\mu 
	\De \right)} - 1 -\frac{1}{m \De} \int_2^{\infty} d\xi
	\frac{\sqrt{\xi^2-4}}{\xi^3}\, (1-e^{-\xi m \De}) - \ln{\left(m
	\De\right)} \cr &&\hspace{2.2cm} +\frac{1}{2} \int_0^1 dx \,
	\ln{\left(1+ \frac{m^2/\mu^2}{x(1-x)} \right)} \Bigg\}\,.
	\label{eq_sing_plat_mass_final1_l3}
\end{eqnarray}
Note that the massless ($m\to0$) and sharp ($\De\to0$)
limits coincide in the $\xi$ integral term and that
\begin{equation}
	\lim_{\eta \to 0} \left( \frac{1}{\eta} \int_2^{\infty} d\xi
	\frac{\sqrt{\xi^2-4}}{\xi^3} (1-e^{-\xi\eta}) + \ln\eta \right)
	= -\gamma.
\end{equation}
In addition, the last term of eq.~(\ref{eq_sing_plat_mass_final1_l3}),
the Feynman parameter integral, vanishes in the massless limit.  So
the leading $\De\to 0$ divergence of ${\cal E}^{(2)}_{\rm 
3D}(\si_{\vert})$ is the same whether or not $\phi$ has a mass.  This
is true for all three geometries we have studied.  Note that the
subleading divergence proportional to $\frac{1}{\De}$ is altered in the case
of a massive scalar boson.

When the mass of $\phi$ is non-zero it is possible to study the two
dimensional case.  This problem has been studied with a Gau{\ss}ian background in
\cite{dens}.  For comparison with the
three-dimensional case, we compute the Casimir energy of  a
single plate (line) in two dimensions with the background of eq.~(\ref{sig2}).

\subsubsection{Two Dimensions}
In two spatial dimensions the vacuum polarization diagram is finite
and requires no renormalization. However, the contribution to the
energy becomes infinite in the sharp limit, as we will now show.  
 The integration in eq.~(\ref{2Dtwopoint}) cannot be performed 
analytically for arbitrary $\Delta$.  However the leading behavior 
as $\Delta\to 0$ can be easily obtained.  The energy per unit length 
is given by 
\be
{\cal E}^{(2)}_{\rm 2D}(\si_{\vert}) =  \frac{\lambda^{2}}{32 \pi}
\ln{(\De m)} + \textrm{terms~finite~for~} \De \to 0.
\label{lim2d}
\ee 
This calculation demonstrates that the contribution to the energy from
the second-order Feynman diagram in two spatial dimensions is
divergent in the limit $\De \to 0$, and this divergence cannot be
renormalized away.  If we try to renormalize the vacuum polarization
graph (by a finite counterterm proportional to $\sigma^2$), the
divergence only gets worse.  The two-point function
\be 
\Pi(|\vek p|) = \frac{1}{4 \pi |\vek p|}
\arctan{\frac{|\vek p|}{2 m}} 
\ee
is renormalized to
\be 
\Pi_{\rm ren}(|\vek p|) = \Pi(|\vek p|) - \Pi(\mu) 
\ee 
where $\mu$ is a renormalization scale and the leading divergence now
is proportional to $1/\Delta$.  A numerical analysis of the vacuum 
polarization energy for a Gau{\ss}ian background peaked on a circle of 
radius $R$ can be found in Ref.~\cite{dens}.   There the energy 
\emph{density} is studied, and the log-divergence that develops as 
$\De\to 0$ can be seen to be concentrated on the emerging surface at 
$R$.

\subsection{Comments}

At this point already some important conclusions may be drawn: first,
the renormalized second order Feynman diagram diverges in the sharp
limit for all the simple geometries we have considered.  Second, since
the Feynman diagram expansion is an expansion in powers of $\lambda$,
the contribution from the second order diagram cannot be canceled by
the other diagrams in the series.  They are of different powers in
$\lambda$.  Third, the diagram has been renormalized, and the
counterterm has been fixed.  Thus, the divergence in the sharp limit
is physical and continuum quantum field theory does not provide a way
to get rid of it.  If one is only interested in forces that do not
require the body to be deformed, one need not be troubled by cutoff
dependence in the energy.  However, if one has to vary the surface
size or geometry as in the case of a stress, the sharp limit gives a
divergent result.  It is already clear at this point that throwing
away divergent terms on an ad hoc basis will eliminate parts of the
physical measurable stress; therefore, in cases where one has to
deform the body to obtain the stress \textit{the sharp limit cannot be
taken}.
\bigskip
\bigskip

\section{The Three-Point Function}

\bigskip

To further clarify the nature of the divergences in the Dirichlet
Casimir energy, we consider the three-point function in a scalar
theory in three dimensions.  No divergences originate from the loop
integrations in the Feynman diagrams at this order (and beyond).  As a
result, we do not need to discuss the subtle issue of regularization
and renormalization and no divergence in the three-point function can
be renormalized away.  It must instead reflect a physical cutoff
dependence generated by the interaction of the fluctuating field with
the background matter.

The three-point function for a fluctuating boson field reduces to
\begin{equation}
	E^{(3)}_{\rm 3D}[\si]=\frac{i\lambda^3}{3T}\int 
        \frac{d^4q_1}{(2\pi)^4} 
	\frac{d^4q_2}{(2\pi)^4}\frac{d^4q_3}{(2\pi)^4}\,
	\frac{\tilde\sigma_4(q_1-q_2)}{q_2^2-m^2+i\eta}\,
	\frac{\tilde\sigma_4(q_2-q_3)}{q_3^2-m^2+i\eta}\,
	\frac{\tilde\sigma_4(q_3-q_1)}{q_1^2-m^2+i\eta}\,.
	\label{Gamma3}
\end{equation}
Here $\tilde\sigma_4(q)=\int d^4x \sigma(x){\rm e}^{iq\cdot x}
=2\pi\de(q^{0})\tilde\si(\vek q)$ is the Fourier transform of the time
independent background field $\sigma(x)$ in four dimensions. 
The factor of $\frac{1}{T}$, where $T$ is a large time interval, converts
the effective action for arbitrary $\si$ to the Casimir energy for
time independent $\si$.

We consider two cases for $\sigma(\vek x)$: First we assume a
spherically symmetric barrier of radius $R$ and width $\De$, given by
eq.~(\ref{sig3}), and second, we consider parallel plates  of
width $\De$  separated by $2L$, given by eq.~(\ref{sig2}). 

As already noted, the integral over the loop momentum in
eq.~(\ref{Gamma3}) converges.  Thus $E^{(3)}_{\rm 3D}[\si]$ can be
evaluated without reference to regularization.   It can be cast into the form
\begin{equation}
E^{(3)}_{\rm 3D}[\si]\propto \int_0^\infty dP f(P)
\label{e3}
\end{equation}
where it is understood that $E^{(3)}_{\rm 3D}[\si]$ is to be replaced
by the energy per unit area, ${\cal E}^{(3)}_{\rm 3D}[\si]$, in the
case of parallel plates.  In eq.~(\ref{e3}) and below we omit
constants of proportionality that are independent of the parameters
($R,L,\De$) characterizing the background field, which are irrelevant
for the present discussion.  For simplicity of notation we also
suppress the dependence of $E$ and $f$ on $R$, $L$, $\De$, $\la$, {\it
etc.\/}

For the spherically symmetric background we find the
integrand 
\begin{eqnarray}
	f_{\circ}(P)&=& P^4\int_0^{\frac{\pi}{2}}d\theta \, \cos
	\theta\, \sin^2\theta \, \sigma_\circ(P\cos\theta)\,
	\sigma_\circ(P\sin\theta)\notag\\ && \hspace{2cm}\times
	\int_{-1}^{1}dz\, \sigma_\circ(P\sqrt{1+z\sin2\theta})\, g(P\cos
	\theta,P\sin\theta,z)\notag
\end{eqnarray}
where
\begin{eqnarray}
	g(p,q,z)&=& \int_0^1 \frac{dx}{\sqrt{4m^2+4x(1-x)q^2+(p+2xqz)^2}}
	\notag\\ && \hspace{2cm}\times \Big\{
	\tanh^{-1}\frac{2xp-p+2xqz}{\sqrt{4m^2+4x(1-x)q^2+(p+2xqz)^2}}
	\notag\\
	&& \hspace{2.4cm}
	-\tanh^{-1}\frac{p+2xqz}{\sqrt{4m^2+4x(1-x)q^2+(p+2xqz)^2}}
	\Big\}\,.
	\label{isphere}
\end{eqnarray}
The integrals in eqs.~(\ref{e3}) and (\ref{isphere}) cannot be
evaluated analytically so we study them numerically.

\begin{figure}
\centerline{ \BoxedEPSF{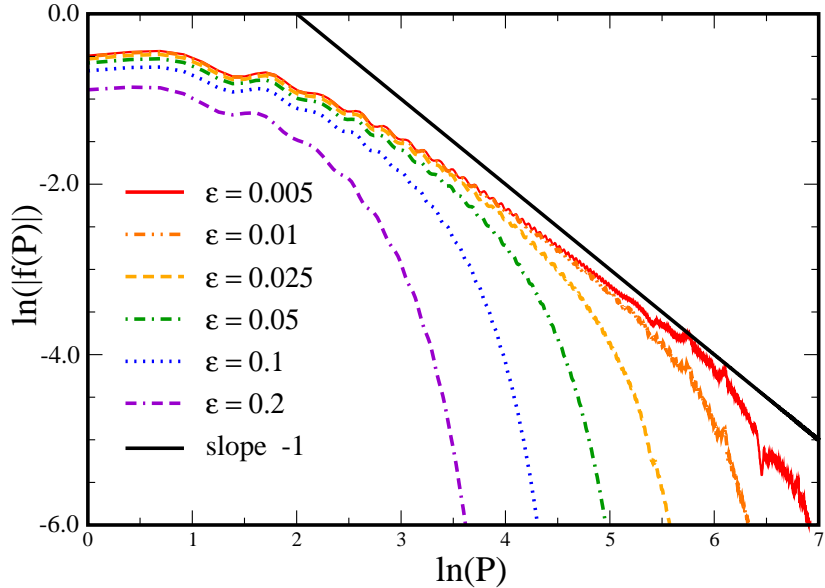 scaled 450}}
\caption{\label{fig_1}\sl The logarithm of the integrand,
eq.~(\ref{isphere}) of the three-point function for the spherical
background, eq.~(\ref{ffsphere}) for various values of the thickness,
$\eps=\De/R$, of the shell.  Energies are measured in $m$, the mass of
the particle in the loop.  For orientation we have also included a
limiting function, $f(P)\sim1/P$, for which $E^{(3)}$ diverges.  Here
the case $R=0.8/m$ is shown.}
\end{figure}

In Fig.~\ref{fig_1} we display a typical example for the integrand of
the spherical problem, eq~(\ref{isphere}).  For a given non-zero
width, $\De$, there is always an intermediate region in which
$f_{\circ}(P)\propto 1/P$.  Beyond that region the integrand drops off
faster than $1/P$ so the integral converges for finite width $\De$. 
As the width decreases, the region in which $f_{\circ}(P)\propto 1/P$
increases and continues to larger values in $P$.  As $\De\to0$, this
behavior persists for arbitrarily large $P$ and therefore the integral
eq~(\ref{e3}) diverges in that limit.  The upper boundary of the
region that exhibits the $1/P$ behavior increases approximately like
$1/\De$.  We can therefore estimate the divergence as a function of
the width,
\begin{equation}
\int_0^\infty dP f_\circ(P)\approx \int_0^{P_0} dP f_\circ(P)
+\int_{P_0}^{1/\De} dP\, \frac{C}{P}
 =C\ln\De + \ldots
\label{estimatediv}
\end{equation}
where  $C$ is a constant and $P_0$ denotes the momentum at
which the $1/P$ behavior sets in.  This logarithmic divergence is
expected from the discussion above: in essence, $1/\De$ serves as a
large momentum cutoff on $f_{\circ}(P)$.

\begin{figure}
\vskip1.0cm \centerline{
\BoxedEPSF{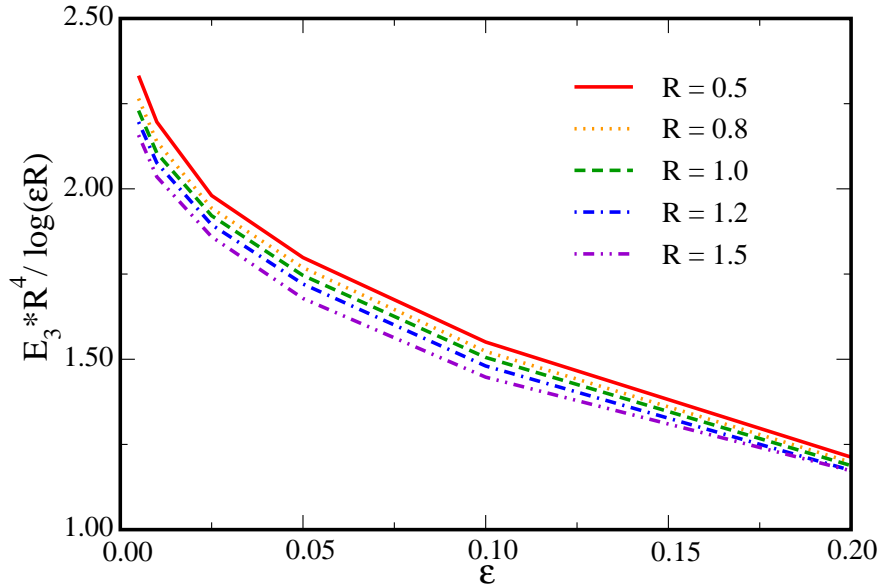 scaled 450}}
\caption{\label{fig_2}\sl The contribution of the three-point function
to the energy as a function of $\eps=\De/R$.  Energies are measured in
units of the mass $m$ of the particle in the loop.  In those units the
coupling constant $\lambda$ is chosen to be 1 and a scale is chosen to
stress the $R$-dependence as well as the logarithmic dependence on
$\De$.}  
\end{figure}

This divergence is studied in Fig.~\ref{fig_2} where we show
$E^{(3)}_{\rm 3D}(\si_{\circ})$ as function of the width parameter for
various values of the position of the spherical barrier.  Over the
considered range, $\De$ varies by about two orders of magnitude.  On
that scale the curves shown Fig.~\ref{fig_2} are approximately
constant and thus verify the logarithmic divergence estimated in
eq.~(\ref{estimatediv}).  So the divergence is not as severe as for
the renormalized two-point function, but it is still present.  As
illustrated in the next section, where we discuss the
$\delta$-function background ({\it ie.\/} $\De\to0$), this is the
expected behavior.  When we restore the dimensional factors we see
that $E^{(3)}_{\rm 3D}(\si_{\circ})\propto \frac{1}{R^{4}}\ln\De$, so
the constant of proportionality is $R$ dependent.  Therefore the
stress ( $-\partial E^{(3)}_{\rm 3D}(\si_{\circ})/\partial R$ ) also
diverges logarithmically as $\De\to0$.  As discussed in the previous
section, the exact form of the $R$ dependence depends on the
normalization we choose for $\si_{\circ}$.  If we switch to the
normalization used for plates where $\int dr\si(r)=1$, then
$E^{(3)}_{\rm 3D}(\si_{\circ})\propto R^2 \ln\De$, so the divergent
contribution to the stress,  and even its sign(!),  depends on
the physical interpretation we place on $\si$, which is by no means
clear.  Note that the $R$ dependence of this divergence differs from
that of the two-point function, so for every choice of normalization
of $\si$ the sum of the two terms diverges.

The case in which the barriers approximate two parallel plates is
simpler because one more integral can be evaluated analytically.  The
corresponding integrand in eq.~(\ref{e3}) then reads,
\begin{align}
	f_{\parallel}(P)&= P\int_0^{2\pi} d\theta\ 
	\Sigma(P\cos\theta,P\sin\theta)\, g(P\cos\theta,P\sin\theta)\  ,\notag \\
	\intertext{where}
	\Sigma(p,q)&= \sigma_\parallel(p)\sigma_\parallel(q) 
	\sigma_\parallel(p+q)\ ,\notag\\
	\sigma_\parallel(q)&= \frac{2}{q\De}\,\cos(qL)\,
	{\rm sin}\left(\frac{q\De}{2}\right)\, ,\notag\\ 
	\intertext{and}
	g(p,q)&= \frac{1}{p}\int_0^1 \frac{dx}
	{\sqrt{4m^2+4x(1-x)q^2+(p+2xq)^2}}\notag \\ &  \hspace{2cm}\times
	\Big\{ \tanh^{-1}\frac{2xp-p+2xq}{\sqrt{4m^2+4x(1-x)q^2+(p+2xq)^2}}\notag \\ & 
	\hspace{2.4cm} -\tanh^{-1}\frac{p+2xq}{\sqrt{4m^2+4x(1-x)q^2+(p+2xq)^2}} \Big\}\,.
	\label{iplates}
\end{align}
In Table~\ref{tab_1} we present numerical results for the energy per
unit area, ${\cal E}^{(3)}_{\rm 3D}(\si_{\parallel})$, which
originates from the three-point function.  Like the sphere, ${\cal
E}^{(3)}_{\rm 3D}(\si_{\parallel})$ diverges like ${\rm ln} \De $ as
$\De\to 0$.  However, this time the divergence is independent of $L$. 
As we found for the two-point function, \emph{the force between
parallel plates due to the three-point function is finite and
unambiguous as $\De \to 0$.}

\begin{table}[t]
\centerline{
\begin{tabular}{l|c c c c c}
$\De$ & $L=0.5$ & $L=0.8$ & $L=1.0$ & $L=1.2$ & $L=1.5$\\
\hline
0.010 & -163.04 & -160.24 & -160.88 & -161.04 & -161.28 \\
0.025 & -122.16 & -119.68 & -119.68 & -119.76 & -120.00 \\
0.050 & -95.36 & -92.64 & -92.16 & -92.08 & -92.24 \\
0.100 & -70.56 & -67.84 & -67.36 & -67.28 & -67.20 \\
0.200 & -48.08 & -44.72 & -45.12 & -45.04 & -44.96 \\
\end{tabular}}
\caption{\label{tab_1}\sl The contribution of the three-point
function to the energy per unit area,  ${\cal E}^{(3)}(\si_{\parallel})$, 
as a function of the width, $\Delta$, and the
distance of the parallel plates, $L$. Units are set by the mass, $m$, of
the particle in the loop.  Constants of proportionality, that do not
vary with either $\De$ or $L$ are omitted.  The variation of 
${\cal E}^{(3)}(\si_{\parallel})$ across the columns of the Table is 
consistent within the accuracy of the calculation, with the divergent 
part of ${\cal E}^{(3)}(\si_{\parallel})$ being independent of $L$.}
\end{table}

To summarize, we have found that the contribution of the three-point
function to the Casimir energy of a sphere or the Casimir energy per
unit area of parallel plates diverges logarithmically in the sharp
limit in three dimensions.  We expect that this divergence, like that
of the two-point function, is the same for all shapes.  This is
because the width parameter essentially plays the role of a cutoff on
the Fourier spectrum of the background, independent of its geometry. 
There is no counterterm in the continuum field theory to cancel this
divergence.  The stress on a sphere suffers the same divergence. 
However the force between parallel plates does not.  Instead it
reaches a finite limit as $\De\to 0$.  In the case of the sphere, the
$R$ dependence of $\si_{\circ}(r)$ could be chosen to render the
divergence independent of $R$.  However no choice can remove the
divergences in both the two and three-point functions.  We conclude
that the Dirichlet-Casimir stress on a sphere in three dimensions
depends in an unavoidable way on the cutoffs that characterize the
material.  It is not possible to compute even the cutoff dependent
term in the stress on a sphere, to say nothing of the finite parts,
because it depends dramatically on what is assumed about the $R$
dependence of $\si_{\circ}(r)$.
 
\section{Vacuum Polarization Energy in a Sharp Background}
\bigskip

In this section we take a different approach to the calculation of the
Casimir energy in a sharp background in three space dimensions. 
Instead of introducing a finite thickness, $\Delta$, as a cutoff, we
start with a $\delta$-function background, $\si^{\ast}$, from the
start and introduce a cutoff, $\Lambda$, on its Fourier spectrum. 
This enables us to compute \emph{all} $n$-point functions for the
parallel plate and spherical geometries and to explore the divergences
in a simpler mathematical context.

We first consider the case of parallel plates.  To begin we impose the 
constraint that $\si^{\ast}$ depends only on $x_{3}$ and is 
symmetric under reflection in this coordinate,
\begin{equation}
	\sigma^{\ast}_{\parallel} (\vek x
	)=\sigma^{\ast}_{\parallel}(x_3)\,,\qquad
	\sigma^{\ast}_{\parallel} (x_3)=\sigma^{\ast}_{\parallel}(-x_3)\,.
	\label{sgplate1}
\end{equation}
The appropriate form of the free Green's function for this problem reads
\begin{equation}
	G_0(x,x^\prime)=\int \frac{d^{\,3}q}{(2\pi)^3}\,
	{\rm e}^{i\tilde{q}\cdot(\tilde{x}-\tilde{x^\prime})}
	\int_{-\infty}^\infty \frac{dp}{2\pi}\sum_{\ell=0,1}
	\frac{\cos(x_3p+\frac{\ell\pi}{2})
	\cos(x^\prime_3p+\frac{\ell\pi}{2})}
	{\tilde{q}^2-p^2-m^2+i\eta}\,.
	\label{greenplates}
\end{equation}
Here $\tilde{v}_\mu=(v_0,v_1,v_2)$ comprises the time component as
well as those spatial components of a four vector, $v_\mu$, orthogonal
to $x_{3}$.  The quantum number $\ell=0,1$ is associated with parity. 
The $n$-point function, $\Gamma^{(n)}(\si^{\ast}_{\parallel})$, is
given by    
\begin{equation}
	\Gamma^{(n)}(\si^{\ast}_{\parallel})=
        \frac{-i\lambda^n}{2n}\int d^4x_1\int
	d^4x_2\ldots\int d^4x_n G(x_1,x_2)\sigma^{\ast}_{\parallel}(\vek x_2
	)G(x_2,x_3)\sigma^{\ast}_{\parallel}(\vek x_3 )\ldots G(x_n,x_1)\,
	\sigma^{\ast}_{\parallel}(\vek x_1 )\,.
	\label{npoint}
\end{equation}
We have already established that
$\Gamma^{(n)}(\si^{\ast}_{\parallel})$ diverges for $n=1,2,$ and $3$. 
They should be regarded as regulated by a cutoff which will be
introduced explicitly below.  For a background of the structure given
in eq~(\ref{sgplate1}) the integrals $\int d^3\tilde x_i$ are
straightforward ($\tilde x =(x^0,x^1,x^2)$).  They yield overall
(infinite) factors of the ``time interval'' and ``area'' of the
plates.  The energy per unit area is obtained by dividing out these
factors.  The order $n$ contribution is (after Wick rotation)
\begin{align}
	{\cal E}_{\rm 3D}^{(n)}(\si^{\ast}_{\parallel})&=  
        -\frac{(-\lambda)^n}{2n}\int
	\frac{d^3\tilde{q}}{(2\pi)^3} \int_{-\infty}^\infty
	\frac{dp_1}{2\pi}\frac{1}{q^2+p_1^2+m^2}\ldots \int_{-\infty}^\infty
	\frac{dp_n}{2\pi}\frac{1}{q^2+p_n^2+m^2} \notag\\ & \hspace{1cm}\times
	\int_{-\infty}^\infty dz_1\ldots dz_n\, 
        t(p_n;z_n,z_1)\sigma^{\ast}_{\parallel}(z_1)
	t(p_n;z_1,z_2)\sigma^{\ast}_{\parallel}(z_2)\ldots 
	t(p_n;z_{n-1},z_n)\sigma^{\ast}_{\parallel}(z_n)\ ,
	\label{enint1}
\end{align}
where
\begin{equation}
	t(p;z,z^\prime) =  \sum_{\ell=0,1}\cos(zp+\frac{\ell\pi}{2})\,
	\cos(z^\prime p+\frac{\ell\pi}{2})\,.
\end{equation}

This expression simplifies dramatically when we assume a
$\delta$-function background
\begin{equation}
	\sigma^{\ast}_{\parallel} (z)= 
          \delta(z-L)+\delta(z+L) 
	\label{delta1}
\end{equation}
The integrals over $z_i$ not only become not only trivial, but
identical, thus
\begin{eqnarray}
{\cal E}_{\rm 3D}^{(n)}(\si^{\ast}_{\parallel},\La)
	&=&-\frac{(-2\lambda)^n}{2n}\int^{\La} \frac{d^{\, 3}q}{(2\pi)^3}
	\sum_{\ell=0,1}\left[\int_{-\infty}^\infty\frac{dp}{2\pi}
	\frac{\cos(pL+\frac{\ell\pi}{2})}{q^2+p^2+m^2}\right]^2
	\cr
	&=&-\frac{1}{2n}\int^{\La} \frac{d^{\, 3}q}{(2\pi)^3}
	\left[\frac{-\lambda}{2\omega(q)}\right]^n
	\left\{(1+{\rm e}^{-2\omega(q)L})^n+
	(1-{\rm e}^{-2\omega(q)L})^n\right\}\,,
	\label{enint2}
\end{eqnarray}
where the cutoff $\La$ appears as an argument for ${\cal E}$ to remind
us that this is a regulated Casimir energy.  We would like to mention
that there is no physical interpretation of this cutoff because it
cannot be directly linked to either the renormalization scale or the
Fourier spectrum of the background field.  This cutoff merely serves
to render eq.~(\ref{enint2}) finite for all $n$.   Simple power
counting indicates that the $n=$ 1, 2, and 3 contributions diverge as
$\La\to\infty$, while, as expected, the contributions of order $n=4$
and higher are finite. 

Keeping the cutoff fixed, all orders are finite and can be summed,
\begin{eqnarray}
	{\cal E}_{\rm 3D} (\si^{\ast}_{\parallel},\La)
	&=&\frac{1}{2}\int^\Lambda \frac{d^{\, 3}q}{(2\pi)^3} \ln \left[
	1+\frac{\lambda}{\omega(q)}
	+\frac{\lambda^2}{4\omega^2(q)}\left(1-{\rm
	e}^{-4\omega(q)L}\right)\right] \cr
	&=&\int_m^{\sqrt{\Lambda^2+m^2}} \frac{dt}{(2\pi)^2}
	t\,\sqrt{t^2-m^2}\ln \left[1+\frac{\lambda}{t}
	+\frac{\lambda^2}{4t^2}\left(1-{\rm e}^{-4tL}\right)\right]\,.
	\label{ensum}
\end{eqnarray}
 It is easy to verify that the divergences of ${\cal E}_{\rm 3D}
(\si^{\ast}_{\parallel},\La)$ as $\La\to\infty$ correspond to those we
analyzed diagram by diagram earlier.  However the divergences are
$L$-independent, so that  we can extract a force per unit area
(pressure) between the two plates,
\begin{equation}
	\mathcal{P}_{\rm 3D}(\si_{\parallel}^{\ast}
	,\La)=-\frac{1}{2}\frac{\partial {\cal E}_{\rm 3D} 
	(\si^{\ast}_{\parallel},\La) }{\partial L}=
        -\frac{\lambda^2}{8\pi^2}
	\int_m^{\sqrt{\Lambda^2+m^2}} dt\, \frac{\sqrt{t^2-m^2}\,{\rm
	e}^{-4tL}} {1+\frac{\lambda}{t}+\frac{\lambda^2}{4t^2}\left(1-{\rm
	e}^{-4tL}\right)}\,.
	\label{force}
\end{equation}
 which is finite as $\Lambda\to\infty$.  Furthermore, it remains
finite in the strong limit in which $\lambda\to\infty$ and the
situation reduces to two Dirichlet plates.  We then obtain the
established result for the force between two Dirichlet plates as a
function of $m$ and $L$\cite{Bordag}.  In the Appendix we show that
the same result is obtained in the framework of the interface
formalism of Ref.~\cite{GJQW}.

Next we consider the analogous calculation for a sharp background with
spherical symmetry, $\sigma^{\ast}_{\circ}(\vek x )=\delta(r-R)/4\pi
R^{2}$.  The starting point is the spherical decomposition of the free
Green's function
\begin{equation}
	G_0(x,x')=\frac{2}{\pi}\int_{-\infty}^{\infty} \frac{d\omega}{2\pi}\,
	{\rm e}^{i\omega\left(x_0-x'_0\right)}\,
	\sum_{\ell m}\,\int_{0}^{\infty} dk k^2\
	j_\ell(kr)j_\ell(kr')\,
	\frac{Y_{\ell m}^{*}(\hat{x})Y_{\ell m}(\hat{x}')}
	{\omega^2-k^2-m^2+i\eta}\,,
	\label{g0sphere}
\end{equation}
where $j_\ell$ are spherical Bessel functions and $Y_{\ell m}$ are
spherical harmonics.  We substitute this decomposition into the
expression for the $n$-point function, eq.~(\ref{npoint}).  Again the
integral over the time and angular coordinates is trivial.  It can be
factorized to extract the $n$-th order contribution to the energy
$E_{\rm 3D}^{(n)}(\si^{\ast}_{\circ})$,  
\begin{eqnarray}
	E_{\rm 3D}^{(n)}(\si^{\ast}_{\circ})& =&\frac{1}{n}
	\left(\frac{-2\lambda}{\pi}\right)^n\, \int_{-\infty}^{\infty}
	\frac{d\omega}{2\pi}\,\sum_{\ell}\left(2\ell+1\right)\,
	\int_{0}^{\infty} \frac{k_1^2 dk_1}{\omega^2+k_1^2+m^2} \ldots
	\int_{0}^{\infty} \frac{k_n^2 dk_n}{\omega^2+k_n^2+m^2} \cr
	&&\times \int_{0}^{\infty} dr_1 r_1^2 j_\ell(k_n
	r_1)\sigma^{\ast}_{\circ}(r_1)j_\ell(k_1 r_1) \ldots
	\int_{0}^{\infty} dr_n r_n^2 j_\ell(k_{n-1}
	r_n)\sigma^{\ast}_{\circ}(r_n)j_\ell(k_n r_n) \qquad
\end{eqnarray}
where the frequency integral has been been Wick--rotated. As
in the case of parallel plates, the contributions for $n=1,2,
\mbox{and} 3$ diverge.  They should be regarded as regulated by a
cutoff to be introduced explicitly below. 

For the $\delta$--function background the momentum integrals can be
done, giving
\begin{eqnarray}
	E_{\rm 3D}^{(n)}(\si^{\ast}_{\circ})&=&
	\frac{1}{n}\left(\frac{-\lambda}{2\pi^2}\right)^n\,
	\int_{-\La}^{\La}
	\frac{d\omega}{2\pi}\,\sum_{\ell}\left(2\ell+1\right)\,
	\left[\int_0^\infty dk k^2\, \frac{j_\ell(k R)^2}
	{\omega^2+k^2+m^2}\right]^n\, \cr
	&=&\frac{2}{n}\left(\frac{-\lambda}{4\pi R}\right)^n\,
	\int_0^{\La} \frac{d\omega}{2\pi}\, \sum_{\ell}
	\left(2\ell+1\right)\,
	\left[I_{\ell+1/2}\left(R\sqrt{\omega^2+m^2}\right)
	K_{\ell+1/2}\left(R\sqrt{\omega^2+m^2}\right)\right]^n\ ,\nonumber
	\\
	\label{ensphere1}
\end{eqnarray}
where $I_\nu$ and $K_\nu$ are modified Bessel functions.  The sum over
$\ell$ is finite for fixed $\omega$, but divergences appear in the
$\omega$ integral for small $n$.  To regulate these divergences we
have introduced a cutoff, $\La$, on the frequency integral.
\begin{figure}
\vskip1.0cm \centerline{ \BoxedEPSF{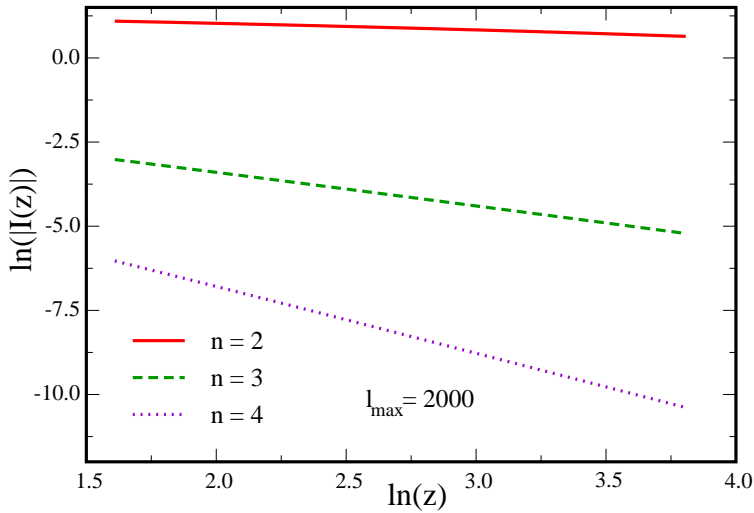 scaled 400}}
\caption{\label{fig_3}\sl The integrand for ${\cal E}^{(n)}$
in eq.~(\protect\ref{ensphere1}) as a function of the integral
variable $z$ as given in eq.~(\protect\ref{integrand}).  The maximal
angular momentum used in this numerical study is $\ell_{\rm
max}=2000$, which is sufficient in the $\omega$ range considered.
}
\end{figure}

The central question is the behavior of the frequency integral after
summing over orbital angular momentum $\ell$.  For $n=1,2$ we expect
divergences because we know that the loop integral in the Feynman
graph diverges.  Since we did not distinguish between loop and
external momenta, Feynman loop integral divergences will show up as
divergences in the frequency integral.  For $n=1$ we know that
$\sum_\ell (2\ell+1) j_\ell(z)^2=1$. After summing $\ell$,
 the $k$ integral diverges linearly, as does the $\omega$ 
integral. These two divergences combine to a quadratically
divergent object as expected for the tadpole graph.  When we renormalize,
the entire $n=1$ term is canceled by a counterterm.  In
figure~\ref{fig_3} we display the integrand,
\begin{equation}
	I_{n}(z)= \sum_{\ell=0}^{\ell_{\rm max}}\left(2\ell+1\right)\,
	\left[I_{\ell+1/2}(z) K_{\ell+1/2}(z) \right]^n\,,\quad
        z=R\sqrt{\omega^2+m^2}\,,
	\label{integrand}
\end{equation}
for $n=2,3,4$ in a log-log plot.  The sum over $\ell$ is performed
numerically up to $\ell_{\rm max}$ large enough to ensure convergence
at the values of $\omega$ considered.  This numerical study
confirms the results we obtained by studying the $n=2$ and
$n=3$ cases in Sections 3 and 4.  For $n=2$ the integrand goes like
$\omega^{-\alpha_{2}}$ with $0<\alpha_{2}<1$.  The best fit to
$\alpha_{2}$ is 0.2064. So the divergences of the frequency integral is
stronger than logarithmic, {\it i. e.\/} stronger than the loop
divergence.  For $n=3$ our best fit to $\alpha_{3}$ is 1.0004,
consistent with the logarithmic divergence that we expect from the
analysis of Section 4.  Finally, for $n=4$ $I_{4}$ behaves like
$1/\omega^2$ (the best fit to $\alpha_{4}$ is 1.9829) yielding a finite
frequency integral.  So do the higher orders in $n$.  This confirms
our earlier results that for a $\delta$-function background  the
singularities are still contained in the lower order Feynman diagrams
and are stronger than those from the loop integrals.  

Finally, we stress that unlike the case of parallel plates, the 
divergences in the case of a sphere  \emph{do depend} on the radius of 
the sphere and therefore afflict the Casimir pressure, making it 
cutoff dependent.

\bigskip
\section{Conclusions}

We have seen that the Casimir energy of a Dirichlet boundary can be
modeled as the limit of the Casimir energy of a background potential,
which can then be calculated using the conventional tools of
renormalized quantum field theory.  Although this energy is finite and
unambiguous for any particular smooth background, the limit of this
process can still diverge, implying that the energy depends in detail
on the properties of the physical material used to impose the boundary
conditions.  For three space dimensions, the Casimir energy has
$\frac{1}{\De}\ln\De$ divergences both for parallel plates and for the
sphere.  These divergences should not be confused with the loop
divergences of Feynman diagrams: they also appear in diagrams that
have no quantum field theory divergences.  In particular, the third
order diagram needs no renormalization but also diverges in the sharp
limit, going like $\ln \De$.

This cutoff dependence may or may not enter into physically measurable
quantities.  The force between rigid bodies is never affected, so our
analysis does not alter standard results such as the force between
parallel plates.  However, it does render meaningless calculations of
stresses, such as the Casimir surface tension of the sphere.  Such
quantities are cutoff dependent, and cannot be defined 
independently of the material properties that determine the cutoff.
 
Our results disagree with those of Ref.~\cite{Milton}, which also considers
low-order Feynman diagrams, but obtains a finite contribution
to the stress in the boundary condition limit.  We disagree with this
calculation on two grounds:  
\begin{itemize}

\item
In Ref.~\cite{Milton} the \emph{background fields} are analytically
continued to fractional dimension.  In dimensional regularization, one
continues loop momenta to fractional dimensions, but background
fields, which are simply complex-valued functions, remain fixed in the
physical dimension (see for example \cite{PS}, p.  662). 
 Analytically continuing the background fields introduces
$\sigma$ dependence into the counterterm coefficients $c_{1}$ and
$c_{2}$ (see eq.~(\ref{ct})).  Counterterm insertions in other diagrams
would then render the theory unrenormalizable.

\item 
In Ref.~\cite{Milton} it is claimed that the contribution of the
two-point function to the Casimir energy is finite for $D=3$ space
dimensions because the divergences appear in the form $\Gamma(1-D/2)$,
which has poles at $D=2$ and $D=4$ but not $D=3$.  However, this
is a familiar misinterpretation of dimensional regularization.  For
example, divergences in QED in $d$ spacetime dimensions appear as
$\Gamma(2-d/2)$, but QED in $4+1$ dimensions is not a finite theory. 
Similarly, the divergences of $\phi^4$ theory come from
$\Gamma(1-d/2)$, but this theory is not finite in $2+1$ or $4+1$
dimensions.  The proper prescription in dimensional regularization is
to use counterterms to  cancel all divergences appearing in
dimensions \emph{less than or equal to} the physical dimension. 

\end{itemize}
If the results described here generalize to the electromagnetic case, as
we expect they do, they would invalidate Boyer's result \cite{Boyer}
that the conducting sphere experiences a cutoff-independent, repulsive
Casimir stress.  For such a result to be correct, it is necessary to
show that it can be obtained as the limit of an underlying smooth,
renormalizable quantum field theory.  Otherwise, the Boyer problem
cannot be studied without reference to material properties, and the
assertion that sphere has repulsive Casimir force is unwarranted.

\paragraph{Acknowledgments}

We gratefully acknowledge discussions with G.~Barton, E.~Farhi and
K.~D.~Olum.  N.~G. is supported in part by the National Science
Foundation (NSF) through the Vermont Experimental Program to Stimulate
Competitive Research (VT-EPSCoR).  R.~L.~J., V.~K., and O.~S. are
supported in part by the U.S.~Department of Energy (D.O.E.) under
cooperative research agreement~\#DF-FC02-94ER40818.  M.~Q., O.~S. and
H.~W. are supported by the Deutsche Forschungsgemeinschaft under
contracts~Qu 137/1-1,~Schr 749/1-1 and We~1254/3-2 respectively.

\bigskip
\section{Appendix:  Interface Formalism and Force between Plates}
 
In this section we will verify that the interface formalism of 
Ref.~\cite{GJQW} applied to a $\delta$-function type background gives
the standard result for the force~\cite{Bordag} between
two parallel Dirichlet plates.

The parallel plate geometry is a standard example of an 
\emph{interface problem}, where $n$ ``trivial'' spatial dimensions 
parallel to the plates are irrelevant for the QFT dynamics and the
equations of motion become effectively $m$-dimensional.
In the particular case of the parallel plates in three dimensions, 
$m=1$ and $n=2$. 

{F}rom the point of view of scattering theory, the interface is
characterized by a separation of the wave functions $\Psi(\vec{x})
\sim \psi(\vec{x}_\perp) e^{i \vec{p}\cdot\vec{x}_\parallel}$ where
$\vec{x} = \vec{x}_\perp + \vec{x}_\parallel$, leading to a continuous
degeneracy of bound and scattering states.  The phase shifts
$\delta(\vec{k})$ will only depend on the $m$-dimensional momenta
$\vec{k}$ in the non-trivial directions, while the integration over
the $n$-dimensional trivial momenta $\vec{p}$ leads to an apparent
\emph{divergence} in the total energy.  In Ref.~\cite{GJQW}, it was
shown that this spurious divergence 
is actually absent due to identities in scattering theory that
generalize Levinson's theorem~\cite{GJQWAnnals}.  It is then possible
to renormalize the Casimir energy per unit area of the interface,
$\mathcal{E} = E/A_{\rm plate}$, with just the standard field theory
counterterms.   For the interesting case $m=1$, $n=2$, the
formula for $\mathcal{E}$ reads ~\cite{GJQWAnnals}
\begin{eqnarray}
\mathcal{E} &=& - \frac{1}{12\pi} \sum\limits_{\ell = \pm} \left[
\int_0^\infty \frac{dk}{\pi} \, h(k)\frac{d}{d k}
\left( \delta_\ell(k) - \delta_\ell^{(1)}(k) - \delta_\ell^{(2)}(k)
\right) + \sum_j h(i \kappa_{j,\ell}) \right] + 
\left[\mathcal{E}_{\rm FD}^{(2)} + \mathcal{E}_{\rm CT}\right]
\nonumber\\*
h(k) &\equiv& (k^2 + \mu^2)^{\frac{3}{2}} - \frac{3}{2} k^2 \mu - \mu^3\,,
\label{MQ1}
\end{eqnarray} 
where $\mu$ is the mass of the fluctuating boson, $\mu^2 -
\kappa_{j,\ell}^2 = \omega_{j,\ell}^2$ are the bound state energies
and the counterterm for the first order diagram $\mathcal{E}_{\rm
FD}^{(1)}$ has been chosen to cancel it completely.  The specific form
of $h(k)$ originates from integrating out the $n$ trivial momenta in
dimensional regularization.   $\delta_{\ell}(k)$ is the phase
shift in the $\ell^{\rm th}$ partial wave, and $\delta_{\ell}^{(1)}$ and
$\delta_{\ell}^{(2)}$ are its first and second Born approximations. Since we
are working in three spatial dimensions, a second Born subtraction has
to be performed in the phase shift $\delta_\ell(k)$, and the
corresponding second order diagram $\mathcal{E}_{\rm FD}^{(2)}$ has to
be added back in.  This diagram is then renormalized by a standard
counterterm $\mathcal{E}_{\rm CT}$.  The final result,
eq.~(\ref{MQ1}), yields a finite expression $\mathcal{E}$ for any
\emph{smooth} background $\sigma(\vec{x}_\perp)$.

The phase shifts in eq.~(\ref{MQ1}) are defined through the $m=1$
dimensional problem of a massive boson scattering off a potential
$\sigma(x)$ on a line, and $\ell = \pm$ denotes the two parity
channels of that problem.  We denote the non-trivial coordinate
perpendicular to the plates by $x$ and write $\vec{x} = \vec{x}_\perp
+ \vec{x}_\parallel = x \,\vec{e}_\perp + y_1 \vec{e}_\parallel^{(1)}
+ y_2 \vec{e}_\parallel^{(2)}$.
  
Before we study the \emph{sharp limit} $\sigma_\parallel(x)
=\delta(x-L)+\delta(x+L)$, it is useful to follow~\cite{GJQW} and
rewrite eq.~(\ref{MQ1}) as an integral over the imaginary axis $k=i t$
using analytic properties of scattering data.  This approach combines
the scattering and bound state contribution to eq.~(\ref{MQ1}) in a
simple and concise formula which is also better from a numerical point
of view~\cite{GJQW}.

The general procedure is explained thoroughly in Ref.~\cite{GJQW}. 
Here, it suffices to mention that the phase shift may be expressed in
terms of the \emph{Jost functions}, $2 \delta_\ell(k) = i
\ln\,F_\ell(k)/F_\ell(-k)$.  Since the integrand in eq.~(\ref{MQ1}) is
odd in $k$, we can extend the integral to $[-\infty,\infty]$ and
deform the contour to a large circle in the upper complex $k$-plane,
where $F_\ell(k)$ is meromorphic.  Owing to the two Born subtractions,
the integrand falls fast enough at $|k|\to\infty$.  In addition, the
residues from simple poles at the bound states $k=i\kappa_j$ cancel
the explicit bound state contribution in eq.~(\ref{MQ1}), leaving only
the discontinuity across the branch cut of the power $(k^2 +
m^2)^{\frac{3}{2}}$ in $h(k)$,
\begin{equation}
\mathcal{E} = \frac{1}{4\pi^2} \int\limits_m^\infty d t\,
t \sqrt{t^2-m^2} \sum_{\ell = \pm}\left[\nu_\ell(t) - \nu_\ell^{(1)}(t)
-\nu_\ell^{(2)}(t)\right] +  \mathcal{E}_{\rm FD}^{(2)} + 
\mathcal{E}_{\rm CT} 
\label{MQ2}
\end{equation}
where $\nu_\ell(t) = \ln F_\ell(k=it)$ is a real function of $t$.

Now we are prepared to study the sharp limit $\sigma(x) =  \delta(x+L)
+ \delta(x-L) $,  which we interpreted as a two-channel problem on the half-line
$x\ge 0$.  Using the well-known Jost functions in the symmetric and
antisymmetric channel, respectively,
\begin{equation}
F_\pm(k) = 1 + i \frac{\lambda}{2k}\left[1 \pm e^{2 i k L}\right]
\qquad \Longrightarrow \qquad
\nu_\pm(t) = \ln\left[1 + \frac{\lambda}{2 t} (1 \pm e^{- 2 L t})\right]
\end{equation}
it is readily seen that the $t$-integral in eq.~(\ref{MQ2})
\emph{diverges} even at finite coupling $\lambda$.  This is the
divergence of the Casimir energy in a sharp background from our
previous discussion.  The $t$-integral in eq.~(\ref{MQ2}) is a compact
way of summing all Feynman diagrams of order three and higher.  As
shown in Section 5 the divergence of this sum in the sharp limit is
\emph{exclusively} due to the third order diagram $\mathcal{E}_{\rm
FD}^{(3)}$.  To see this, we subtract one more Born term in
eq.~(\ref{MQ2}) and add back in the corresponding Feynman diagram
$\mathcal{E}_{\rm FD}^{(3)}$.  Then the $t$-integral (now the sum of
all diagrams of order four and higher) will be finite, while the third
order diagram (which would be finite for a smooth background)
diverges \footnote{Note that it is  (far)  simpler to compute
this diagram by (Born) expanding $\nu_\ell(t)$ in powers of the
coupling $\lambda$ and inserting in eq.~(\ref{MQ2}), than to compute 
the Feynman diagram directly.},  {\it cf.\/} eq.~(\ref{enint2}),
 
\begin{equation}
\mathcal{E}_{\rm FD}^{(3)} =  \frac{\lambda^3}{48 \pi^2} \int\limits_m^\infty
dt\, \frac{\sqrt{t^2-m^2}}{t^2} \left[1 + 3 e^{-4 L t}\right].
\label{MQ3}
\end{equation}
There is no field theoretic counterterm to renormalize this diagram and
thus the divergence of the Casimir energy implies a real
physical cutoff dependence.

In the present case, however, this cutoff-dependence has no physical
consequences since the logarithmic divergence in eq.~(\ref{MQ3}) comes
from the first term in the brackets, which turns out to be independent
of the distance $d=2L$ between the plates.  The same holds for the
second order diagram $\mathcal{E}_{\rm FD}^{(2)}$.  As a consequence,
the physically measurable \emph{pressure}
$\mathcal{P}=-d\mathcal{E}/d(2L)$ on the plates is finite even though
the total Casimir energy diverges.  This is again just our main
statement about the cutoff independence of the Casimir force between
\emph{rigid bodies}.

After a little algebra, the subsequent \emph{strong coupling limit}
$\lambda\to\infty$ yields a finite result\footnote{Note that the
finite contributions to the force \emph{cancel} between the Feynman
diagram and the corresponding Born subtraction for both the $2^{\rm
nd}$ and $3^{\rm rd}$ order contributions.  This is important since
those terms, which are of order $\lambda^2$ and $\lambda^3$,
respectively, would otherwise prevent us from taking the strong
coupling limit $\lambda\to\infty$.  Incidentally, this observation
implies that we could have computed the \emph{force} more easily by
making only \emph{one} Born subtraction and adding nothing back in.}
for the Casimir pressure between two parallel plates in three
dimensions.
\begin{equation}
P(d) = -\frac{\partial \mathcal{E}}{\partial (2L)} = -\frac{C(m d)}{d^4}\,,
\qquad\qquad C(s) \equiv \frac{1}{2\pi^2} \int\limits_s^\infty d\tau\,
\frac{\tau^2 \sqrt{\tau^2 - s^2}}{e^{2 \tau} - 1}\,,
\end{equation}
where $d=2L$ is the distance between the plates. This result
can also be derived from eq~(\ref{force}) with $\lambda\to\infty$
and $\Lambda\to\infty$.
The integral $C(s)$ becomes particularly simple in the \emph{massless} 
case, where we recover the standard result
\begin{equation}
P(d) = -\frac{\pi^2}{480\,d^4}\,,\qquad\qquad m=0\,. 
\end{equation}

\end{document}